\documentclass[lettersize,journal]{IEEEtran}
\usepackage{amsmath,amsfonts}
\usepackage{array}
\usepackage[caption=false,font=normalsize,labelfont=sf,textfont=sf]{subfig}
\usepackage{textcomp}
\usepackage{stfloats}
\usepackage{url}
\usepackage{verbatim}
\usepackage{graphicx}
\usepackage{cite}
\usepackage{diagbox}
\usepackage{etoolbox}
\usepackage{multirow}
\usepackage{bbding}
\usepackage{amssymb}
\usepackage[hidelinks]{hyperref} 
\usepackage[ruled,vlined]{algorithm2e}
\SetCommentSty{textit}
\SetKwComment{Comment}{// }{}
\SetKwProg{Fn}{Function}{:}{}
\SetKw{Parameters}{Parameters:}
\SetKw{Initialization}{Initialization:}
\SetKw{Step}{Step}

\begin{document}

  \title{GPR Hierarchical Synergistic Framework for Multi-Access MPQUIC in SAGINs}

  \author{Hanjian Liu, Jinsong Gui
  \thanks{TXXXXXXXXXXXXXXXXXXXXXXXX.}
  \thanks{Manuscript received XX XX, XXXXX; revised XX XX, XXXX.}}

  \markboth{Journal of \LaTeX\ Class Files,~Vol.~XX, No.~X, XX~XXXX}%
  {Shell \MakeLowercase{\textit{et al.}}: A Sample Article Using IEEEtran.cls for IEEE Journals}

  \IEEEpubid{0000--0000/00\$00.00~\copyright~XXXX IEEE}

  \maketitle

  \begin{abstract}
    The deployment of Multipath QUIC (MPQUIC) in Unmanned Aerial Vehicle (UAV)-assisted Space-Air-Ground Integrated Networks (SAGINs) is severely hampered by the out-of-order (OFO) packet delivery problem. Frequent stream handovers, high mobility, and massive multi-access contention in these networks introduce severe transport-layer challenges. Existing solutions typically isolate multipath scheduling from congestion control, which leads to suboptimal performance and transient congestion in highly dynamic environments. To overcome these limitations, this paper proposes the GPR Hierarchical Synergistic Framework, representing the first joint optimization of multipath scheduling and congestion control for multi-access MPQUIC in SAGINs. Our framework introduces the GradNorm Probabilistic Self-Predictive (GPASP) module to forecast latent states and filter task-irrelevant information in high-dimensional, noisy observation spaces. Furthermore, we develop a Proactive Handover-Aware Congestion Control (PHACC) algorithm that leverages neural network-driven decisions to proactively distinguish handover-induced packet losses from actual network congestion. To address decision-making lag caused by neural network inference latency, a Neural-network Preference Estimation (NNPE) algorithm is designed for highly efficient, real-time scheduling. Extensive ns-3 simulations demonstrate that the proposed framework significantly outperforms state-of-the-art baselines, achieving substantial goodput improvements and a marked reduction in OFO degrees.
  \end{abstract}


  \section{Introduction}
    \IEEEPARstart{W}{ith} the rapid emergence of the sixth-generation (6G) communication era, the Space-Air-Ground Integrated Network (SAGIN) has been envisioned as a revolutionary paradigm to provide ubiquitous, high-capacity, and globally seamless connectivity \cite{lang2025joint}. By synergistically combining terrestrial infrastructures with Non-Terrestrial Networks (NTNs) such as Low Earth Orbit satellites (LEO SATs) and Unmanned Aerial Vehicles (UAVs), SAGINs can significantly expand coverage to remote and disaster-stricken areas \cite{fan2025cooperative}. In this architecture, UAVs serve as highly agile aerial relays or base stations, offering line-of-sight (LoS) links and flexible deployment to support massive User Equipments (UEs) \cite{liu2025uav}. However, the inherent characteristics of SAGINs—such as the high mobility of both UAVs and LEO SATs, frequent handovers, intermittent coverage, and highly heterogeneous link properties—pose severe challenges to ensuring reliable and low-latency end-to-end data transmission.

    Despite these architectural advantages, the highly dynamic topology of UAV-assisted SAGINs introduces severe communication challenges at the transport layer. The continuous, high-speed movement of both UAVs and LEO SATs results in fluctuating link qualities, asymmetric path delays, and frequent handovers \cite{huang2025deep}. Traditional single-path transport protocols, such as TCP and QUIC, suffer from severe performance degradation and connection interruptions in such volatile environments. Nevertheless, multipath QUIC (MPQUIC) offers a novel solution to address these challenges. As a user-space extension of QUIC, MPQUIC inherits features like 0 round-trip time (RTT) connection establishment, built-in encryption, and connection migration via Connection IDs (CIDs), while enabling the simultaneous aggregation of multiple network paths. This intrinsically positions MPQUIC as a promising candidate for improving transmission reliability and throughput in mobile non-terrestrial networks \cite{kimura2025evaluating}.

    However, the deployment of MPQUIC in UAV-assisted SAGINs is severely hampered by the out-of-order (OFO) packet delivery problem. Due to significant disparities in bandwidth, latency, and packet loss rate (PLR) among parallel paths in SAGINs, packets scheduled across different paths frequently arrive at the receiver out of sequence, causing head-of-line blocking, buffer overflow, and throughput degradation \cite{liu2025energy}. Existing literature primarily addresses the OFO problem by proposing either novel multipath schedulers to predict path delays \cite{huang2025quiccourier} or new congestion control (CC) algorithms to dynamically adjust sending windows \cite{wang2025ccolia}. However, these solutions are predominantly designed for relatively stable terrestrial networks or idealized parameter variations \cite{qian2026parameter, suleman2026simulated}. When applied to SAGINs, they struggle to adapt to the extreme heterogeneity and transient congestion triggered by UAV mobility and SAT-UAV handovers. Furthermore, current studies tend to focus on either scheduling or congestion control in isolation, failing to coordinate the intrinsic coupling between the two. In highly dynamic SAGIN environments, the mismatch between aggressive congestion control and conservative scheduling often leads to suboptimal performance and severe network congestion.

    \IEEEpubidadjcol

    Moreover, existing research predominantly focuses on single-user multipath transmission scenarios. In dense 6G networks, however, multiple UEs concurrently compete for the limited and time-varying bandwidth of UAV relays and SAT backhauls. Optimizing MPQUIC in such multi-access scenarios introduces a high-dimensional, noisy observation space and exacerbates the partial observability problem, rendering traditional heuristics and standard Reinforcement Learning (RL) approaches brittle and sample-inefficient \cite{wang2025resilient}. Additionally, in highly dynamic SAGINs, neural network inference latency can induce decision-making lag, resulting in ineffective actions or even system performance degradation.

    To dress these challenges, this paper presents the first study on MPQUIC optimization in UAV-assisted SAGINs. We perform a joint optimization of multipath scheduling and congestion control to maximize overall system performance. Specifically, we propose a novel GPR Hierarchical Synergistic Framework tailored for multi-access MPQUIC in SAGINs. The main contributions of this paper are summarized as follows:

    \begin{enumerate}
    \item \textbf{Novel Joint Optimization Framework:} To our best knowledge, this is the first study on joint optimization of multipath scheduling and congestion control for multi-access MPQUIC in UAV-assisted SAGINs, and we establish a systematic model quantifying the out-of-order (OFO) degree and overall throughput.
    \item \textbf{GradNorm Probabilistic Self-Predictive:} We design a GPASP module to tackle the high-dimensional, noisy observation space and partial observability issues inherent in multi-access SAGINs. Unlike traditional observation-predictive methods, GPASP utilizes self-predictive representations to forecast latent states, thereby filtering task-irrelevant information. Furthermore, an adaptive GradNorm mechanism is integrated to dynamically balance RL and auxiliary tasks, ensuring stable convergence and sample efficiency.
    \item \textbf{Proactive Handover-Aware Congestion Control:} We develop a PHACC algorithm that leverages neural network-driven scheduling decisions to achieve proactive handover awareness. Specifically, we propose an Exponentially Decaying Boost Slow Start (EDBSS) mechanism to smooth the congestion window growth, and a multi-metric packet loss differentiation mechanism to distinguish handover-induced loss from real congestion. This approach prevents misjudgment-induced throughput fluctuations and ensures robust congestion control during frequent SAT-UAV handovers.
    \item \textbf{Real-time Inference and Resilience Enhancement:} To mitigate the decision-making lag caused by neural network inference latency, we propose a Neural-network Preference Estimation (NNPE) algorithm. Drawing on cognitive psychology, NNPE derives suboptimal scheduling strategies efficiently without exhaustive real-time inference. Additionally, a Resilient Hierarchical Reward Monitor (RHRM) is implemented to dynamically evaluate system performance and trigger hierarchical recovery logic, ensuring the system's robustness against environmental stochasticity and distributional shifts.
    \item \textbf{Comprehensive Performance Evaluation:} We extend the ns-3 simulation platform to support MPQUIC in SAGIN scenarios. Extensive simulation results demonstrate that the proposed framework significantly outperforms state-of-the-art baselines, achieving substantial improvements in goodput and a marked reduction in OFO degrees under highly dynamic and heterogeneous network conditions.
    \end{enumerate}

  \section{Related Work}
    \label{sec:related_work}
    The deployment of Multipath QUIC (MPQUIC) in Space-Air-Ground Integrated Networks (SAGINs) lies at the intersection of non-terrestrial networking, multipath packet scheduling, and intelligent congestion control. In this section, we review the existing literature surrounding these domains and highlight the unresolved challenges that motivate our work.

    \subsection{Multipath Transport in Non-Terrestrial Networks}

    To overcome the coverage limitations of terrestrial infrastructures and the high volatility of mobile wireless channels, SAGINs have been widely recognized as a pivotal architecture for 6G \cite{lang2025joint}. However, the high mobility of Low Earth Orbit (LEO) SATs and Unmanned Aerial Vehicles (UAVs) introduces frequent handovers, intermittent coverage, and asymmetric path delays. To enhance transmission reliability in such environments, multipath transmission and scheduling strategies have been extensively studied. For instance, Zhang \textit{et al.} proposed a prediction-based MPQUIC scheduler (PBMS) to support real-time applications over ultra-dense LEO SAT networks \cite{zhang2026pbms}. Broadening the scope to network-level traffic coordination, Li \textit{et al.} developed a hierarchical deep reinforcement learning (DRL) and Stackelberg game framework to optimize multipath differential routing and traffic scheduling in ultra-dense LEO constellations \cite{li2025efficient}. While these studies confirm the potential of multipath mechanisms in Non-Terrestrial Networks (NTNs), they primarily rely on the predictable orbital parameters of SATs to model topology variations and focus predominantly on space-segment optimizations. Consequently, they fail to adequately address the extreme topological shifts, transient congestion, and severe multi-access contention introduced by highly agile UAVs in dense, integrated SAGIN environments.

    \subsection{Multipath Scheduling and OFO Mitigation}

    A significant challenge in multipath transport over heterogeneous networks is the Out-of-Order (OFO) packet delivery problem, which causes head-of-line blocking, buffer overflow, and severe throughput degradation. Traditional scheduling algorithms, such as minRTT, ECF, and BLEST, rely on simplistic heuristics that perform poorly under the dynamic latency fluctuations typical of mobile networks \cite{xing2023hbes}. Consequently, researchers have increasingly turned to learning-based solutions. Han \textit{et al.} introduced MARS, a multi-agent DRL scheduler that incorporates OFO queue size feedback to adapt to dynamic environments \cite{han2024mars}. Xu \textit{et al.} proposed DOFMS, a Double Deep Q-Network (DDQN) framework aimed at balancing throughput and OFO rates in mobile heterogeneous networks \cite{xu2024dofms}. Similarly, approaches like LooM \cite{luo2025looM} and MAMS \cite{yang2024mams} attempt to mitigate OFO delivery by predicting path conditions and introducing blocking delays. Despite these advancements, existing learning-based schedulers typically operate independently of the underlying congestion control mechanisms. In highly dynamic SAGINs, this isolation leads to a critical mismatch between aggressive congestion window scaling and conservative scheduling, ultimately inducing transient congestion.

    \subsection{Intelligent Congestion Control}

    Coupled congestion control (CC) algorithms (e.g., LIA, OLIA, BALIA) synchronize subflows to satisfy fairness constraints but are famously sluggish in adapting to rapid path capacity variations \cite{liu2025incc, xu2026afcc}. To address this, Wang \textit{et al.} proposed CC-OLIA, which utilizes packet loss classification to avoid drastic window reductions during independent loss events in mobile networks \cite{wang2025ccolia}. For SAT environments, Yang \textit{et al.} developed a Mobility-Aware Congestion Control (MACO) algorithm that utilizes a Bandwidth-Delay Product (BDP)-inspired quick start \cite{yang2024maco}. Furthermore, DRL has been applied to design advanced CC algorithms. Yu \textit{et al.} proposed MPLibra, a hybrid framework that bridges classic MPTCP mechanisms with DRL to ensure TCP-friendliness and convergence \cite{yu2024mplibra}, while Wang \textit{et al.} introduced TransAL-CC, utilizing Transformer encoders and asynchronous RL to extract fine-grained temporal dependencies \cite{wang2025transal}. However, in the context of UAV-assisted SAGINs, packet losses are frequently triggered by SAT-UAV handovers rather than actual buffer overflows. Existing intelligent CC schemes lack proactive handover awareness and multi-metric loss differentiation, leading to severe throughput fluctuations caused by signal misjudgments.

    \subsection{DRL in Multi-Access SAGINs}

      DRL is increasingly utilized to solve complex, multi-objective optimization problems in SAGINs, such as resilient massive access and power allocation \cite{wang2025resilient} and two-timescale intelligent task offloading \cite{sun2026knowledge}. Nevertheless, scaling DRL to multi-access MPQUIC scenarios remains challenging. When multiple User Equipments (UEs) concurrently compete for the time-varying bandwidth of UAV relays and SAT backhauls, the state space becomes exceptionally high-dimensional and noisy. Standard DRL algorithms suffer from severe partial observability and sample inefficiency under these conditions. Furthermore, the inherent inference latency of complex neural networks induces decision-making lag, rendering real-time packet scheduling ineffective in fast-fading SAGIN channels. 

    In contrast to the aforementioned works, this paper proposes the first joint optimization of multipath scheduling and congestion control tailored for multi-access MPQUIC in SAGINs. By introducing the GPR Hierarchical Synergistic Framework, filtering task-irrelevant noise via the GPASP module, and mitigating inference lag with the NNPE algorithm, our approach intrinsically resolves the disconnect between scheduling and congestion control via the PHACC algorithm, thereby providing resilient, high-throughput, and handover-aware communications in 6G architectures.

  \section{System Model}

\begin{figure}[!ht]
      \centering
      \includegraphics[width=0.48\textwidth]{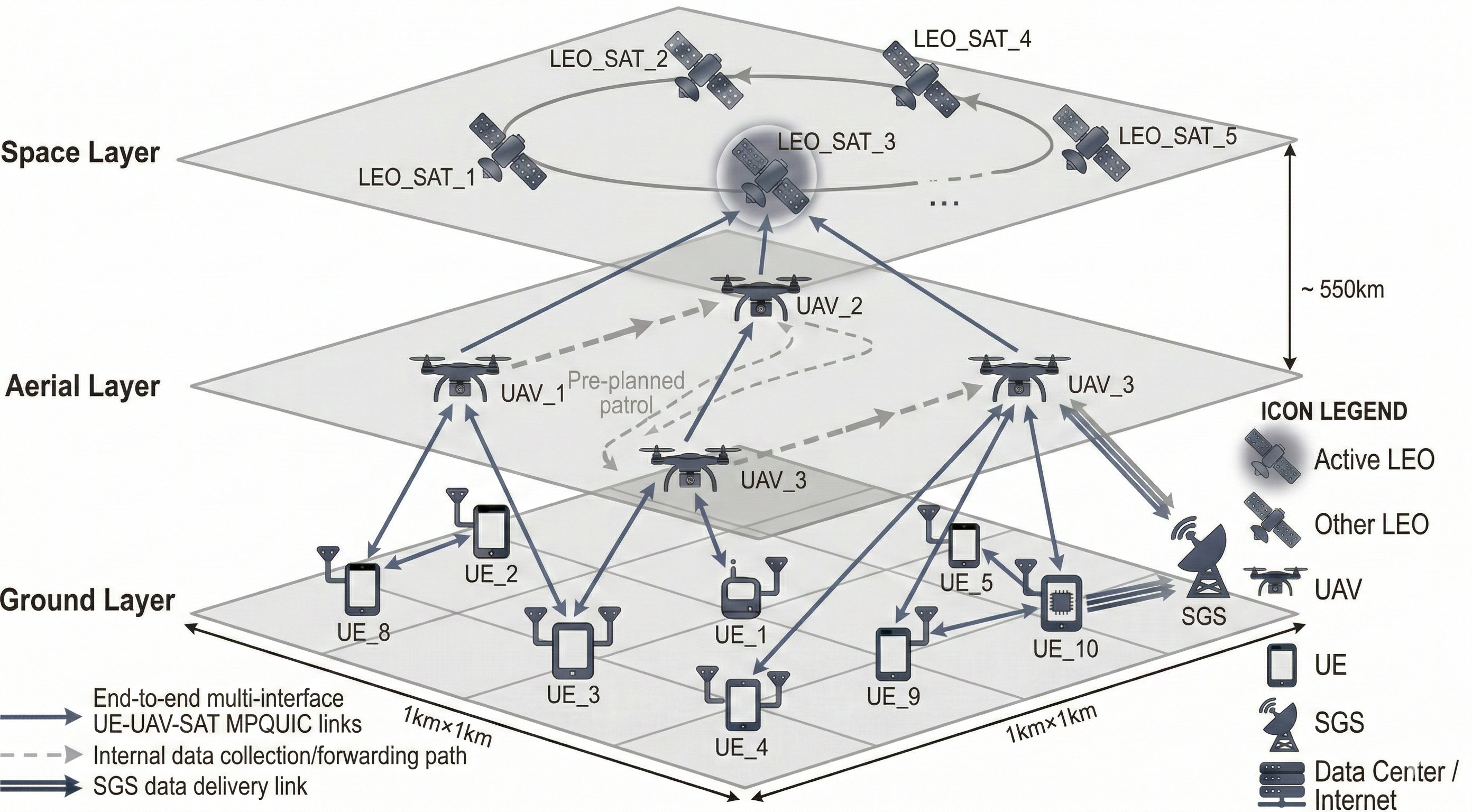}
      \caption{MPQUIC in UAV-assisted SAGIN.}
      \label{fig:NET}
    \end{figure}

    Fig. \ref{fig:NET} illustrates the communication scenario of the Space-Air-Ground Integrated Network (SAGIN) incorporating Unmanned Aerial Vehicles (UAVs), which comprises randomly distributed ground User Equipments (UEs), mission-specific UAVs flying opportunistically over these sensing nodes, and Low Earth Orbit (LEO) SATs passing periodically above them. Without loss of generality, assume that the MPQUIC-enabled UEs have multiple interfaces to connect with accessible UAVs, \(L\) LEO SATs in one orbit take turns to serve the target area,  and the UAVs transmit all collected data packets to the same closest SATs. The SAT ground station is primarily responsible for delivering SAT-collected data to remote data centers but cannot provide high-quality data transmission services for a large number of UEs. The service period \(T\) of each LEO SAT is equally divided into \(N_T\) time slots with a slot length of \(\tau\), i.e., \(T = N_T\tau\). In each time slot, UEs select one UE-UAV-SAT MPQUIC link to transmit data packets of the current flow.

    The set of GTs, UAVs, and the closest SATs are represented by \(\mathcal{N} = \{1, \dots, n, \dots, N\}\), \(\mathcal{M} = \{1, \dots, m, \dots, M\}\), and \(l\). As UAVs serve as relay nodes in UE-UAV-SAT links, each UAV corresponds to one MPQUIC link from the perspective of a UE. Thus, for simplicity and intuitiveness, all links of each UE are directly denoted by \(\{1, \dots, m, \dots, M\}\). Let \((p^n_1,\dots,p^n_i,\dots,p^n_{x_n})\) denote the set of all \(x_n\) packets from UE-\(n\) to SAT-\(l\), and \((x^n_1,\dots,x^n_i,\dots,x^n_{x_n})\) denote their arrival order. Therefore, OFO degree can be defined as: 
    \begin{align}
      & f(p^n_i)=
      \begin{cases}
      x^n_i-x^n_{i+1}, & x^n_i>x^n_{i+1} \\
      0, & x^n_i\leq x^n_{i+1}  
      \end{cases} \\
      & F^{ofo}=\left(\sum_{i=1}^{x_n}f\left(p^n_i\right)\right)/x_n. \label{eq:ofo_degree}
    \end{align}

    Accordingly, overall throughput is defined as:
    \begin{align} \label{eq:overall_throughput}
      & D(t)=\sum_{n=1}^{N}x_n.
    \end{align}

  \section{Problem and MDP Formulation}
    
    Our primary objective is to determine a policy \(\pi\) that coordinates the per-slot multipath scheduling of UEs and effectively performs congestion control, with the aim of maximizing goodput and minimizing the OFO degree during each SAT service period. This can be formulated as follows:
    \begin{align}
      \nonumber
      & \underset{\mathbf{\pi}}{\text{max}} \ \omega_1D^G-\omega_2F^{OFO},\\
        &\text{s. t.}~(\ref{eq:ofo_degree}), (\ref{eq:overall_throughput}), \nonumber\\
        &C1:\mathbf{q}_n^U(t) \in [0, L_E]^2, \forall n \in \mathcal{N}, \nonumber\\
        &C2:\sum_{n=1}^N s_n^m(t) \leq Y^U, \forall m \in \mathcal{M}, \nonumber\\
        &C3:d_{m,l}(t) \leq d_{max}, \forall m \in \mathcal{M}, l \in \mathcal{L}, \nonumber\\
    \end{align}
    where \(\omega_1\) and \(\omega_2\) denote the weights of the objective function. Constraint \(C1\) ensures that UEs are within the service area of size \(L_E \times L_E\); constraint \(C2\) limits the number of concurrent services supported by each UAV; and constraint \(C3\) guarantees that all UAVs are within the visibility range of the SAT.

  \section{GPR Hierarchical Synergistic Framework}
    
\begin{figure}[!ht]
      \centering
      \includegraphics[width=0.48\textwidth]{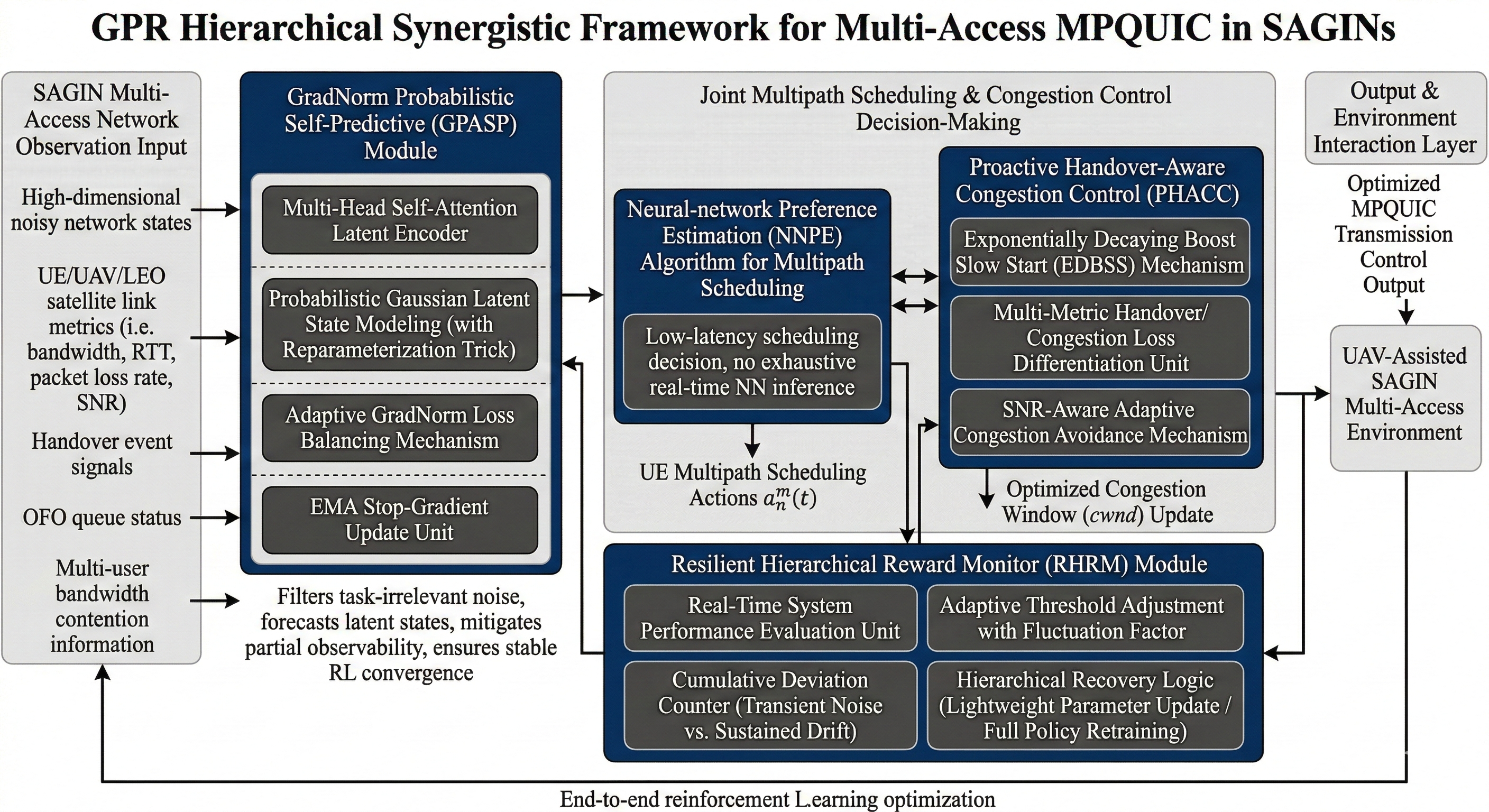}
      \caption{GPR Hierarchical Synergistic Framework.}
      \label{fig:GRP}
    \end{figure}

    In this section, GPR Hierarchical Synergistic Framework is proposed to address the issues of multipath scheduling and congestion control, as shown in Fig. \ref{fig:GRP}. This module comprises three core components: GradNorm Probabilistic Self-Predictive (GPASP), Proactive Handover-Aware Congestion Control (PHACC), and Resilient Hierarchical Reward Monitor (RHRM).

    \subsection{State Encoder}

    \begin{algorithm}[!t]
    \caption{GradNorm Probabilistic Self-Predictive} \label{alg:GPSP}
    \LinesNumbered
    \SetCommentSty{footnotesize} %
    \KwIn{self-attention encoder $f_{\phi}$ and target encoder $f_{\overline{\phi}}$, policy $\pi_{\nu}$ and old policy $\pi_{\overline{\nu}}$, value network $V_{\omega}$, latent transition model $g_{\theta}$, EMA decay $e$, adaptation rate $\eta$, initial loss coefficient $\lambda$, learning rate $\alpha$, smoothing factor $\beta$.}
    \KwOut{updated parameters $\phi, \theta, \nu, \omega$.}

    \For{$t=1$ to $N_{st}$}{

    $z_t \leftarrow f_r(f_{\phi}(h_t))$, $(\overline{\mu}_{t+1},\overline{\sigma}_{t+1}^2) \leftarrow f_{\overline{\phi}}(h_{t+1})$,
    $\overline{z}_{t+1} \leftarrow \overline{\mu}_{t+1}$;\\

    $\delta_t \leftarrow r_t + \gamma V_{\omega}(\overline{z}_{t+1}) - V_{\omega}(z_t)$;\\
    $\hat{A}_t \leftarrow \delta_t + \gamma \lambda_{GAE} \hat{A}_{t+1}$;\\
    $\hat{V}_t \leftarrow \hat{A}_t + V_{\omega}(z_t)$;\\

    $a_t \sim \pi_{\overline{\nu}}(\cdot|z_t), \quad
    r_t \leftarrow \frac{\pi_{\nu}(a_t|z_t)}{\pi_{\overline{\nu}}(a_t|z_t)}$;\\
    $\mathcal L_{\pi} \leftarrow -\min\left( r_t \hat{A}_t, \text{clip}(r_t,1-\epsilon,1+\epsilon)\hat{A}_t \right)$;\\
    $\mathcal L_V \leftarrow \frac{1}{2}( V_{\omega}(z_t) - \hat{V}_t )^2$;\\
    $\mathcal L_{\text{RL}} \leftarrow \mathcal L_{\pi} + \mathcal L_V - \mathbb{E}_{a_t\sim\pi_{\nu}}\left[\log\pi_{\nu}\left(a_t\mid z_{t}\right)\right]$;\\

    $(\hat{\mu}_{t+1},\hat{\sigma}_{t+1}^2) \leftarrow g_{\theta}(z_t,a_t)$;\\
    $\displaystyle
    \mathcal L_{\text{aux}} \leftarrow
    \frac{1}{2}\sum_i
    \Big(
    \log\frac{\hat{\sigma}_{t+1,i}^2}{\overline{\sigma}_{t+1,i}^2}
    + \frac{\overline{\sigma}_{t+1,i}^2 + (\overline{\mu}_{t+1,i}-\hat{\mu}_{t+1,i})^2}
    {\hat{\sigma}_{t+1,i}^2}
    -1
    \Big);$\\

    $G_{\text{RL}} \leftarrow \|\nabla_{\phi}\mathcal L_{\text{RL}}\|_2,\;
    G_{\text{aux}} \leftarrow \|\nabla_{\phi}\mathcal L_{\text{aux}}\|_2$;\\
    $\lambda \leftarrow \lambda \times
    \left( \frac{G_{\text{RL}}}{G_{\text{aux}}+\varepsilon} \right)^{\eta}$,\quad
    $\lambda \leftarrow \text{clip}(\lambda_{\min},\lambda,\lambda_{\max})$;\\

    $\mathcal L_{\text{total}} \leftarrow \mathcal L_{\text{RL}} + \lambda \mathcal L_{\text{aux}}$;\\
    $[\phi,\nu,\omega,\theta] \leftarrow [\phi,\nu,\omega,\theta] - \alpha \nabla \mathcal L_{\text{total}}$;\\
    $[\overline{\phi},\overline{\nu},\overline{\omega}] \leftarrow \beta [\overline{\phi},\overline{\nu},\overline{\omega}] + (1-\beta)[\phi,\nu,\omega]$;
    }
    \end{algorithm}

     RL for multi-access multipath transmission in Space-Air-Ground Integrated Networks (SAGIN) is impeded by high-dimensional, noisy observations and partial observability, rendering traditional algorithms brittle and sample-inefficient. In such contexts, given the performance degradation of observation-predictive methods which directly predict the next state, we adopt self-predictive representations ($\phi_L$) to forecast latent states rather than raw observations, thereby filtering task-irrelevant information and ensuring stability \cite{nietzke2024bridging}. Building on this, we propose the GradNorm Probabilistic Self-Predictive (GPSP), which integrates: a self-attention probabilistic latent transition encoder to capture long-range temporal correlations and model uncertainty in highly dynamic environments (Lines 2--3, Algorithm \ref{alg:GPSP}); an adaptive GradNorm mechanism to dynamically balance RL and auxiliary tasks for stable convergence (Lines 13--15, Algorithm \ref{alg:GPSP}); and stop-gradient with exponential moving average (EMA) updates to prevent representational collapse (Lines 16--17, Algorithm \ref{alg:GPSP}).

    The  multi-head attention encoder \(f_{\phi}\) to obtain the parameters \((\mu_t,\sigma_t^2)\) of a latent Gaussian distribution is:
    
    \begin{equation}
      \begin{split}
        f_{\phi}(h_t) &= \mathrm{MLP}\!\Bigg(\sum_{i=1}^L \beta_i\,
        \Bigg(\mathrm{||}_{k=1}^K\Big[W_O^kW_V^k h_{t,j} \\
        &\quad \sum_{j=1}^L \mathrm{softmax}\Big(\frac{(W_Q^k h_{t,i})^\top (W_K^k h_{t,j})}{\sqrt{d_k}}\Big)\Big]\Bigg)\Bigg), \\
      \end{split}
    \end{equation}
    where $h_t\in\mathbb{R}^{L\times d_h}$ denotes the input sequence of $L$ historical feature vectors, and $\parallel$ denotes concatenation. For head $k$, learnable linear maps $(W_Q^k,W_K^k,W_V^k)\in\mathbb{R}^{d_k}$ compute scaled-dot-product attention across positions, with $W_O^k$ projecting each head's aggregated output. Scalar weights $\beta_i = \frac{\exp(q^\top h_{t,i})}{\sum_{r=1}^L\exp(q^\top h_{t,r})}$ are generated by a learnable query $q$ over position-wise representations; concatenated and projected head outputs are pooled via $\beta_i$, then fed into an MLP to yield latent mean $\mu_t$ and log-variance $\log\sigma_t^2$.

    When the encoder generates a probability distribution over latent states, sampling directly from this distribution constitutes a stochastic process that hinders the direct backpropagation of gradients. Therefore, the Reparameterization Trick is employed to separate the stochasticity from the parameters, rendering the sampling process differentiable. The computation is illustrated as follows:
    \begin{equation} \label{eq:Reparameterization_Trick}
      f_r(\mu_t,\sigma_t^2)= \mu_t + \exp(0.5\log\sigma_t^2)\odot\epsilon,\epsilon\sim\mathcal N(0,1).
    \end{equation}

    \subsection{Decision Maker}

      \begin{algorithm}[!t]
        \caption{Proactive Handover-Aware Congestion Control.}\label{alg:PHACC}
        \LinesNumbered
        \KwIn{window queue \(\mathcal{W}_n^m\), available bandwidth \(TP_n^m(t)\), predicted bandwidth \(C_n^m\)\((t-1)\), measured RTT \(\tau_n^m\)\((t-1)\), estimated propagation \(D_n^m(t-1)\), scaling factor \(\gamma\), standard deviation \(\Delta_R\), \(\rho\), \(\forall n \in \mathcal{N}\), \(\forall m \in \mathcal{M}\)\;}
        \KwOut{cwnd \(w_n^m(t)\).}
        \If{\(w_n^m(t-1)<SST\)}{
          \If{subflow \(F_n^m\) is created or reconnected}{
            \If{\(\mathcal{W}_n^m \neq \emptyset\)}{
              \(\hat{w}_n^m \leftarrow EMA \{\mathcal{W}_n^m\}\)\;
            }
            \ElseIf{\(\exists \mathcal{W}_{n,i} \neq \emptyset\)}{
              \(\hat{w}_n^m \leftarrow 0\), \(cnt \leftarrow 0\)\;
              \ForEach{$i\in \mathcal{M}(i \neq m)$}{
                \(\hat{w}_n^m \leftarrow \hat{w}_n^m + EMA\{\mathcal{W}_{n,i}\}\)\;
                \(cnt \leftarrow cnt + 1\)\;
              }
              \(\hat{w}_n^m \leftarrow \frac{\hat{w}_n^m}{cnt}\)\;
            }
              \Else{
                \(\hat{w}_n^m \leftarrow 4 MSS\)\;
              }
            \(w_n^m(t) \leftarrow \min \left\{ B_n^m, \hat{w}_n^m \right\}\)\;
          }
          \Else{ 
          Increase \(w_n^m(t)\) according to Eqs. \eqref{eq:cwnd_update}-\eqref{eq:boost}\;}
        }
        \Else{
          $Con\_1 \Leftarrow \mathbf{1}\{a_n^m(t) \neq a_n^m(t-1) \vee d_n^m(t) > d_{th} \vee d_{m,l}(t) > d'_{th}\}$\;
          $Con\_2 \Leftarrow \mathbf{1}\{TP_n^m(t) < C_n^m(t-1)\}$\;
          $Con\_3 \Leftarrow \mathbf{1}\{\tau_n^m(t) > D_n^m(t-1) + \Delta_R\}$\;
          \If{packet loss $\land$ $Con\_1$}{
            \If{$Con\_2$ $\land$ $Con\_3$}{
              $w_n^m(t) \leftarrow \frac{w_n^m(t-1)}{2}$\;
            }
            \Else{ 
              $w_n^m(t) \leftarrow \gamma w_n^m(t-1)$\;
            }
          }
          \Else{
            $T_n^m(t) \leftarrow \max_{t \in \{t_0, t_1, \ldots, t\}} \tau_n^m(t-1)$\;
            $\lambda_n^m(t) \leftarrow \sigma \frac{SNR_n^m}{SNR_{max}} + (1-\sigma) \lambda_n^m(t-1)$\;
            $w_n^m(t) \leftarrow w_n^m(t-1) + \lambda_n^m(t) \frac{3\max_{m \in \mathcal{M}} \left(\frac{w_n^m(t-1)}{\tau_n^m(t-1)}\right)^2\sqrt{T_n^m(t)}}{2\tau_n^m \left(\sum_{i \in \mathcal{M}} \frac{w_{n,i}(t-1)}{\tau_{n,i}(t-1)}\right)^{\frac{5}{2}}}$\;
          }

      }
      \Return{\(w_n^m(t)\)}\;
      \end{algorithm}

      By feeding the state information of UEs, UAVs, and Sats, along with potential noise and interference, into the aforementioned GPSP module, we obtain all UE multipath scheduling action denoted as \(a_n^m(t),\forall n \in \mathcal{N}, \forall m \in \mathcal{M}\). This action specifies that the data packet of UE-$(n)$ is scheduled to the $m$-th transmission path at time slot $t$. 

      However, frequent multipath scheduling switches and the inherent dynamics of links pose severe challenges to MPQUIC congestion control in the SAGIN architecture. Traditional algorithms frequently suffer from performance degradation due to congestion signal misjudgment when addressing packet loss and topology variations triggered by scheduling and handover. Different from the Mobility-Aware Congestion control (MACO) algorithm \cite{yang2024maco} that estimates BDP mainly relying on physical motion laws, our multipath scheduling and handover decisions are dominated by neural networks, which endows the system with a priori handover awareness. In view of this, we propose the Proactive Handover-Aware Congestion Control (PHACC) algorithm, which fully leverages scheduling decision information provided by neural networks, integrates historical link information and link quality perception, and implements targeted optimizations in both the slow-start and congestion avoidance phases. 
      
      The PHACC algorithm comprises three mechanisms: 1) a historical information-driven cross-flow slow start mechanism that aggregates cross-flow historical window information ($\mathcal{W}_n^m$) via exponential moving average (EMA) to optimize the initial window, thereby shortening the slow-start period (Lines 1--15, Algorithm \ref{alg:PHACC}); 2) a packet loss differentiation mechanism based on multi-metric congestion detection that combines throughput and delay variations ($Con_2, Con_3$) with DRL scheduling decisions and distance thresholds ($Con_1$) to distinguish handover-induced loss from real congestion, applying conservative window reduction ($\gamma w_n^m$) for the former and halving for the latter to prevent misjudgment-induced fluctuations (Lines 17--24, Algorithm \ref{alg:PHACC}); and 3) an SNR-aware adaptive congestion avoidance mechanism that introduces a dynamic coefficient $\lambda$ based on SNR and load balancing degree $\rho$, replacing the square-root growth function to dynamically adapt to channel quality for efficient capacity convergence (Lines 26--28, Algorithm \ref{alg:PHACC}). The calculation methods for the input parameters $TP_n^m(t)$, $C_n^m(t-1)$, $\tau_n^m(t-1)$, and $D_n^m(t-1)$ of Algorithm \ref{alg:PHACC} are detailed in \cite{yang2024maco}.

      Furthermore, to achieve a smoother and congestion-aware growth of the congestion window (cwnd) during the slow start phase, we propose the Exponentially Decaying Boost Slow Start (EDBSS). This algorithm updates the cwnd at each RTT in accordance with the following equations:
      
      \begin{gather}
        w_n^m(t+1) = w_n^m(t)\min\Big(1 + \xi\big(w_n^m(t)\big) \zeta\big(w_n^m(t)\big),\; M_{max}\Big), \label{eq:cwnd_update}\\
        \xi(w_n^m(t)) = \frac{1}{1 + \exp\left(a\left(\frac{w_n^m(t)}{SST}-b\right)\right)},  \label{eq:base_gf}\\
        \zeta(w_n^m(t)) = 1 + \gamma \exp\left(-\frac{w_n^m(t)}{\varrho}\right), \label{eq:boost}
      \end{gather}
      where \(\xi\big(w_n^m(t)\big)\) is an S-shaped gain (sigmoid) that smooths the transition from exponential to linear increase, ensuring the increment rate decreases progressively as \(\mathrm{cwnd}\) approaches Slow Start Threshold (SST); typical sigmoid parameters are \(a=10\) and \(b=0.5\). \(\gamma\) modulates initial \(\mathrm{cwnd}\) amplification: for small \(\mathrm{cwnd}\) the factor \((1+\gamma)\) accelerates early growth beyond conventional slow-start. Increasing \(\gamma\) improves responsiveness but may induce transient congestion if excessive. \(\zeta\big(w_n^m(t)\big)\) denotes the early-stage exponential boost: for small \(\mathrm{cwnd}\) it satisfies \(\mathrm{boost}\approx 1+\gamma\) and decays exponentially with \(\mathrm{cwnd}\) under control of \(\varrho\). \(\varrho\) governs the decay of the exponential boost with cwnd: larger \(\varrho\) prolongs the boost and delays convergence to the SST, whereas smaller \(\varrho\) yields faster decay and more conservative behaviour. Finally, the upper bound \(M_{\max}\) (e.g., \(M_{\max}=2\)) caps the per-RTT multiplicative increase to preserve stability and avoid buffer overflow.

      This slow-start algorithm incorporates an S-shaped attenuation function and an exponential decay term to achieve progressively controlled slow start. In the early phase ($cwnd \ll SST$), both $\xi(w_n^m(t))$ and $\zeta(w_n^m(t))$ approach their maximum values, enabling accelerated window growth that is comparable to or slightly faster than conventional slow start. As cwnd increases, $\xi(w_n^m(t))$ gradually decreases following a logistic function, while $\zeta(w_n^m(t))$ decays exponentially, ensuring a smooth reduction in the growth rate. When cwnd approaches SST, the growth rate nears 1, thereby eliminating oscillations and reducing the risk of overshooting congestion thresholds. EDBSS thus delivers smooth, stable, and self-regulated cwnd growth behavior, preserving fast ramp-up in the early stage while ensuring congestion safety near the transition to congestion avoidance. 

    \subsection{Reward Monitor}

     \begin{algorithm}[!t]
      \caption{Resilient Hierarchical Reward Monitor.}\label{alg:reward_monitor}
      \LinesNumbered
      \KwIn{current reward $rwd$, reward window $rwd\_win$, smoothing factor $\alpha_0$, deviation multiple $mul_0$, trigger threshold $thr_0$, minimum sample size $\hat{n}$, sliding window size $\widehat{W}$, reference update factor $\hat{\beta}$, stability threshold $\hat{\delta}$, severity threshold $\hat{\lambda}$}
      \If{$rwd\_win$ is empty}{
        $\mathbf{do\ initialization}$: $srwd \leftarrow rwd,\ dev \leftarrow 0.5 rwd,\ cnt \leftarrow 0,\ rwd\_win\leftarrow\emptyset,\ arwd\leftarrow 0,\ \hat{\alpha}\leftarrow\alpha_0,\ mul\leftarrow mul_0,\ thr\leftarrow thr_0$\;
      }
      $rwd\_win \leftarrow rwd\_win \cup \{ rwd \}$\;
      $arwd \leftarrow \frac{arwd(|rwd\_win|-1) + rwd}{|rwd\_win|}$\;
      $dev \leftarrow (1-\hat{\alpha})dev + \hat{\alpha} |rwd-srwd|$\;
      \If{$|rwd\_win| < \hat{n}$}{
        \Return\;
      }
      \If{$|rwd\_win| \geq \widehat{W}$}{
        $rwd\_win \leftarrow rwd\_win.pop()$\;
      }
      $F \leftarrow \min\left(\frac{dev}{srwd}, 1\right)$, 
      $\hat{\alpha} \leftarrow \alpha_0(1+F)$,
      $mul \leftarrow mul_0(1-F)$,
      $thr \leftarrow thr_0(1+F)$ \;
      \If{$\frac{|arwd-srwd|}{srwd} < \hat{\delta}$}{
        $srwd \leftarrow (1-\hat{\beta})srwd + \hat{\beta} arwd$ \;
      }
      \If{$rwd > arwd+mul \times dev$}{
        \If{$cnt < 0$}{$cnt \leftarrow 0$\;}
        $cnt \leftarrow cnt + 1$\;
      }
      \If{$rwd < arwd - mul \times dev$}{
        \If{$cnt > 0$}{$cnt \leftarrow 0$\;}
        $cnt \leftarrow cnt - 1$\;
      }
      \If{$|cnt| > thr$}{
        \eIf{$F < \lambda$}{
          Update  $\widehat{\theta}_n$ via Eq. \eqref{eq:estimator}\;
        }{
          Retrain neural network parameters via Algorithm \eqref{alg:GPSP}
        }
      }
      \end{algorithm}

      In highly dynamic Space-Air-Ground Integrated Networks (SAGIN), the inherent inference latency of Neural Network (NN)-based schedulers often leads to decision staleness, degrading real-time multipath scheduling and may induce transient congestion. To mitigate this, we propose a Neural Network Preference Estimation (NNPE) algorithm for multipath scheduling, exploiting an empirical insight from cognitive psychology: human response times are inversely related to preference strength \cite{li2024enhancing}. Specifically, we model the binary ACK feedback as the choice variable $c_i^n$ and the packet transmission delay as the response time $\tilde{t}_i^n$. Then by leveraging equation \eqref{eq:estimator}, NNPE efficiently estimate the preference weight vector of each UE without requiring exhaustive real-time neural inference. The details of NNPE are provided subsequently.

      Each packet transmission generates a binary feedback signal
      \begin{equation}
      c_i^n =
      \begin{cases}
      +1, & \text{if ACK of \(p_i^n\)  is successfully received}, \\
      -1, & \text{if \(p_i^n\) is lost (timeout)}.
      \end{cases}
      \end{equation}

      Let $t_i^n$ denote the measured RTT upon successful reception of the ACK for $p_i^n$, and $T_{\max}$ represent the predefined timeout threshold. We define the response time $\tilde{t}_i^n$ as
      \begin{equation}
      \tilde{t}_i^n =
      \begin{cases}
      t_i^n, & \text{if ACK of \(p_i^n\) is received}, \\
      T_{\max}, & \text{if \(p_i^n\) is lost (timeout)}.
      \end{cases}
      \end{equation}

      The preference weight vector of UE-$n$ is calculated as follows:
      \begin{equation}
      \label{eq:estimator}
      \widehat{\theta}_n
      =
      \Bigg(
      \sum_{i=1}^{x_n}S^{a_{n,i}}_{n,i} (S^{a_{n,i}}_{n,i})^\mathsf{T} \Bigg)^{-1}
      \sum_{i=1}^{x_n} \left(S^{a_{n,i}}_{n,i} \frac{c_i^n}{\tilde{t}_i^n} \right),
      \end{equation}
      where $S^{a_{n,i}}_{n,i}$ denotes the feature vector of the corresponding link when UE-$n$ transmits packet $p_i^n$ by executing action $a_{n,i}$. The operator \((\cdot)^\mathsf{T}\) eliminates sample-distribution bias. The ratio \(\displaystyle c_i^n / \tilde{t}_i^n\) measures the preference-strength rate: large positive values indicate strong preference, large negative values indicate strong aversion, and values near zero denote inconsistent (weak) preference. Finally, the optimal multipath scheduling strategy for UE-$n$ is obtained via \(\arg \max_{m \in \mathcal{M}} S_{n,i}^m\widehat{\theta}_n\).

      By  considering the link characteristics of communication networks and the decision-making rules of neural networks comprehensively, this method enables the derivation of sub-optimal multipath scheduling strategies that maximize multipath scheduling efficiency while strictly guaranteeing transmission reliability.
      
      During the inference phase, deployed DRL agents are susceptible to environmental stochasticity and distributional shifts, rendering static policy parameters suboptimal \cite{xing2023anonline}. To address this, we implement a Resilient Hierarchical Reward Monitor mechanism (Algorithm \ref{alg:reward_monitor}) to dynamically evaluate system performance and trigger necessary adaptations. Specifically, this mechanism integrates: adaptive threshold adjustment driven by a fluctuation factor $F$ to calibrate sensitivity based on real-time reward volatility (Lines 8--10, Algorithm \ref{alg:reward_monitor}); a cumulative counter strategy to distinguish sustained performance drifts from transient noise (Lines 13--20, Algorithm \ref{alg:reward_monitor}); and a hierarchical recovery logic where mild deviations invoke lightweight parameter updates to compensate for gradual environmental drift, while severe persistent anomalies trigger full PPO policy retraining to restore optimality under large distribution shifts. (Lines 21--25, Algorithm \ref{alg:reward_monitor}).

\section{Simulation Experiments}

    The work in \cite{shu2023a} extended the ns-3 QUIC module \cite{quic-ns-3} to MPQUIC in accordance with the IETF draft \cite{liu2022multipath}. Building upon the publicly available implementation \cite{mpquic-ns3}, we aligned the project with the latest ns-3 CMake architecture and employed \texttt{cppyy} to bind the MPQUIC C++ source code into a Python module. This facilitates a Python-centric experimental environment, enabling seamless integration with advanced DRL algorithms.

    \subsection{Simulation Setting}

      In the simulation scenario, nine UEs are randomly distributed within a $1\,\text{km} \times 1\,\text{km}$ area, while four UAVs execute area patrol and target detection missions along pre-planned trajectories. Consistent with Starlink specifications \cite{astronomy2023starlink}, we simulate a Low Earth Orbit (LEO) constellation at an altitude of 550 km.

      Given that our approach enhances system transmission performance through both multipath scheduling and congestion control, we categorize the comparative schemes into two groups. For multipath scheduling, we evaluate the random scheduling baseline, the widely used MinRTT and Round Robin (RR), the novel contextual multi-armed bandit (CMAB)-based Peekaboo \cite{wu2020peekaboo}, the recently proposed QoS-driven Contextual MAB (QC-MAB) \cite{yang2025qos}, and our proposed GPASP. For congestion control, we compare our proposed PHACC against, ablation scheme PHACC w/o GPASP, the widely adopted MPQUIC algorithm OLIA \cite{khalili2013mptcp}, and the novel MACO \cite{yang2024maco}, which is tailored for mobile scenarios. Meanwhile, to ensure a fair comparison and verify the effectiveness of our scheme by controlling variables, all path scheduling algorithms utilize our PHACC for congestion control, while all congestion control algorithms employ our GPASP for path scheduling.

    \subsection{Multipath Scheduling Performance}

      \textit{1) Network capacity comparison:} Fig.\ref{fig:p1}\subref{fig:p1a} illustrates the average throughput convergence over 1000 training episodes. Traditional heuristic algorithms (RR, MinRTT) exhibit consistently low throughput (\(<3\)~Mbps and \(<5\)~Mbps, respectively), as static heuristics fail to capture high-dimensional spatio-temporal dependencies and rapidly fluctuating link characteristics in UAV-assisted SAGINs, lacking mechanisms to anticipate transient link quality variations from UAV mobility and SAT handovers, leading to mismatched scheduling. Learning-based schedulers (Peekaboo, QC-MAB) improve performance via contextual feedback and adaptive exploration but plateau early (\(\simeq6\)~Mbps and \(\simeq8\)~Mbps, respectively), due to partial observability and high noise in SAGINs degrading context abstraction and arm selection effectiveness. In contrast, the proposed GPASP framework achieves the highest throughput, stabilizing at approximately 12 Mbps. This superior performance arises from its probabilistic self-predictive latent encoder—which filters task-irrelevant noise and preserves long-range temporal correlations to improve state predictability—and its adaptive GradNorm mechanism, which balances RL and auxiliary tasks, mitigates gradient conflicts, and accelerates convergence. These designs collectively enable stable utilization of transient high-capacity UAV-SAT paths.

      {\begingroup
      \let\origsize\normalsize
      \renewcommand{\normalsize}{\scriptsize}%
      \begin{figure}[!ht]
        \centering
        \subfloat[Throughput convergence]{\includegraphics[width=0.24\textwidth]{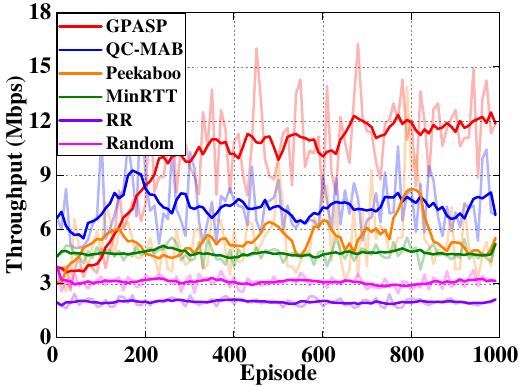}%
        \label{fig:p1a}}
        \subfloat[Delay-Jitter trade-off with PDR]{\includegraphics[width=0.24\textwidth]{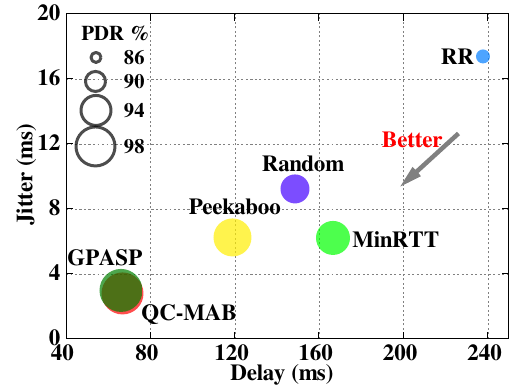}%
        \label{fig:p1b}}
        \hfil
        \subfloat[PLR and OFO rate]{\includegraphics[width=0.24\textwidth]{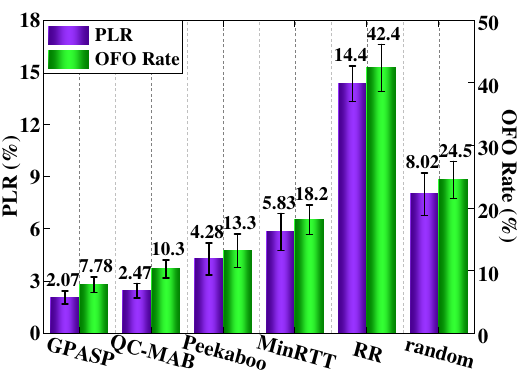}%
        \label{fig:p1c}}
        \subfloat[OFO degree distribution]{\includegraphics[width=0.24\textwidth]{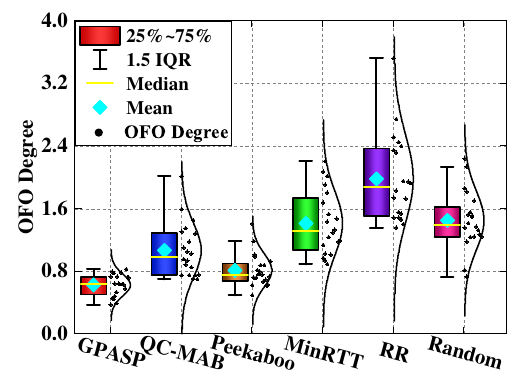}%
        \label{fig:p1d}}
        \caption{Multipath scheduling performance comparison of different schemes.}
        \label{fig:p1}
      \end{figure}
      \endgroup}

      \textit{2) Transmission efficiency comparison:} Fig.\ref{fig:p1}\subref{fig:p1b} depicts the delay-jitter-packet delivery ratio (PDR) trade-off. GPASP and QC-MAB cluster in the bottom-left region with the largest bubbles, indicating comparable low latency, bounded jitter, and high reliability. Despite distinct internal mechanisms, QC-MAB and GPASP converge to similar delay-jitter-PDR performance via functionally aligned decision principles: QC-MAB explicitly embeds delay/reliability constraints into arm selection (biasing toward low-risk paths), while GPASP infer optimal decisions via ACK timing/response latency. Thus, GPASP reduces packet waiting time and scheduling-induced jitter, achieving QoS comparable to QC-MAB in this metric. However, GPASP outperforms QC-MAB in global performance (evidenced by throughput, PLR, and OFO results) via enhanced long-term prediction and tighter scheduler-congestion control coupling—features not explicitly addressed by QC-MAB.

      \textit{3) Network stability comparison:} Fig.\ref{fig:p1}\subref{fig:p1c} compares PLR and Out-of-Order (OFO) rate across schedulers. RR and MinRTT have high PLRs (14.4\% and 5.83\%, respectively) due to rigid rules routing traffic to congested/degrading links, especially during UAV–SAT handovers. Peekaboo and QC-MAB improve reliability (PLRs: 4.28\% and 2.47\%), while GPASP achieves the lowest PLR (2.07\%) via probabilistic latent transition modeling that captures long-range temporal dependencies and anticipates link degradation, reallocating traffic proactively to mitigate queue overflow and packet drops—critical in SAGINs where loss is dominated by mobility-induced link dynamics. 
      
      \textit{4) OFO degree comparison:} Fig.\ref{fig:p1}\subref{fig:p1d} presents OFO degree distribution. RR performs worst (highest median, largest variance) by neglecting path heterogeneity. MinRTT moderately reduces OFO degree but is vulnerable to outdated RTT estimates that fail to track rapid channel fluctuations.QC-MAB and Peekaboo alleviate packet reordering but still have broader interquartile ranges and higher OFO rates than GPASP. Notably, GPASP achieves the lowest median OFO degree, tightest interquartile range, and minimum absolute OFO rate (7.78\%), enabled by two synergistic mechanisms: the probabilistic latent encoder forecasts path delays to minimize cross-path delivery variance, and NNPE mitigates inference lag to synchronize packet dispatching with dynamic link states, reducing reordering/buffer blocking for stable in-order delivery in SAGINs.

      Notably, RR exhibits inferior overall performance to Random, primarily because it uses a rigid, deterministic heuristic that distributes packets blindly across all paths regardless of substantial differences in bandwidth, latency, and packet loss. In contrast, while also non-adaptive, random scheduling prevents the systematic synchronization of packets to poor-quality paths caused by strict rotation.

    \subsection{Congestion Control Performance}

      {\begingroup
      \let\origsize\normalsize
      \renewcommand{\normalsize}{\scriptsize}%
      \begin{figure}[!ht]
        \centering
        \subfloat[Throughput convergence]{\includegraphics[width=0.24\textwidth]{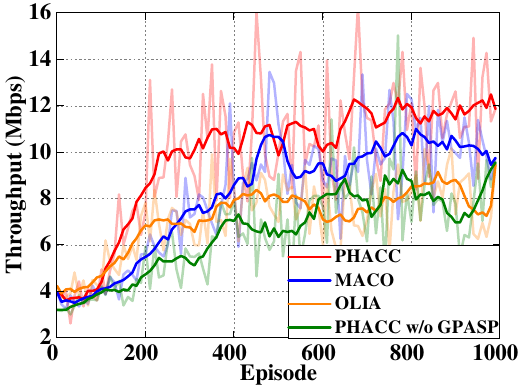}%
        \label{fig:p2a}}
        \subfloat[Delay-Jitter trade-off with PDR]{\includegraphics[width=0.24\textwidth]{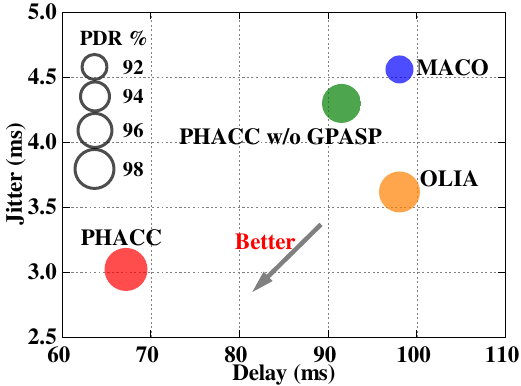}%
        \label{fig:p2b}}
        \hfil
        \subfloat[PLR and OFO rate]{\includegraphics[width=0.24\textwidth]{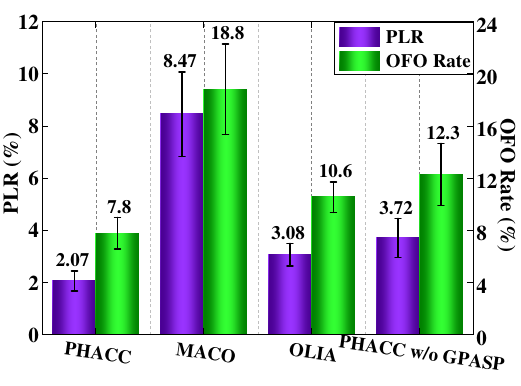}%
        \label{fig:p2c}}
        \subfloat[OFO degree distribution]{\includegraphics[width=0.24\textwidth]{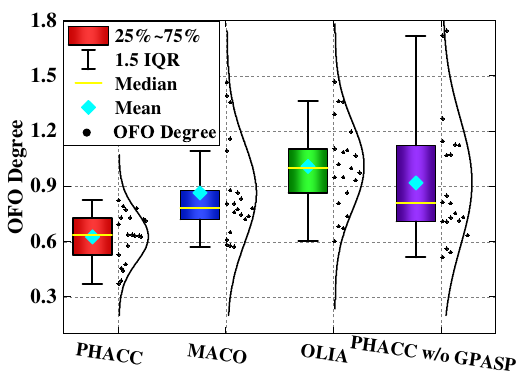}%
        \label{fig:p2d}}
        \caption{Congestion control performance comparison of different schemes.}
        \label{fig:p2}
      \end{figure}
      \endgroup}

      \textit{1) Network capacity comparison:} Fig.\ref{fig:p2}\subref{fig:p2a} presents throughput convergence over training/evaluation runs. OLIA achieves the lowest average throughput ($\approx$8–9 Mbps) owing to conservative subflow synchronization and slow probing; MACO improves to $\approx$10.5 Mbps via mobility-informed BDP estimates, enabling faster probing in most cases. PHACC attains the highest steady throughput ($\approx$12 Mbps), while its ablation (PHACC w/o GPASP) yields lower throughput. Mechanistically, PHACC’s advantage stems from two complementary design choices: (1) Exponentially Decaying Boost Slow Start (EDBSS) uses an S-shaped gain and exponential decay term to balance aggressive early cwnd growth—amplifying probing when cwnd $\ll$ SST while attenuating growth near the estimated SST to reduce overshoot; (2) PHACC leverages scheduler-provided priors (GPASP) and historical cross-flow window statistics (EMA of $W^m_n$) to set safer initial windows and adjust probing aggressiveness for handover anticipation. In contrast, MACO relies on physics-based BDP estimation and AKF correction, which is less effective under heavy multi-access contention and non-access bottlenecks. The ablation confirms scheduler–CC coupling enhances capacity utilization: without GPASP’s noise-filtered latent predictions, PHACC’s priors are less reliable, leading to more conservative or oscillatory cwnd control depending on the scenario.

      \textit{2) Transmission efficiency comparison:} Fig.\ref{fig:p2}\subref{fig:p2b} plots delay (x), jitter (y), and PDR (bubble size). PHACC clusters in the desirable bottom-left (low delay/low jitter) with the largest bubbles (highest PDR), while OLIA and MACO locate in higher-delay regions. This arises from PHACC’s multi-metric loss differentiation (Con1–Con3 in Algorithm 2): when packet loss aligns with handover indicators (e.g., sudden throughput drop, RTT deviation, scheduling switch), PHACC adopts conservative cwnd reduction (factor $\gamma$) instead of aggressive halving for confirmed congestion. This selective response reduces abrupt transmission stalls and queuing-induced delay spikes, tightening jitter and improving PDR. Ablation results show these benefits are amplified when the congestion controller receives reliable, anticipatory scheduling signals from GPASP.

      \textit{3) Network stability comparison:} Fig.\ref{fig:p2}\subref{fig:p2c} compares PLR and OFO rates. OLIA and PHACC w/o GPASP show moderate PLRs (3.08\% and 3.72\%), while PHACC reduces PLR to 2.07\%. PHACC’s lower PLR stems not from being ``less aggressive'' but from accurately distinguishing mobility-induced losses from congestion via short-term throughput/RTT measurements, scheduling-derived handover signals, and SNR-aware weighting in congestion-avoidance. This enables targeted cwnd reductions to avoid unnecessary post-handover rate drops while responding firmly to genuine congestion. The result (lower PLR with high throughput) reflects PHACC’s ability to probe capacity aggressively when safe and back off gently for mobility-related losses. MACO’s mobility-aware BDP estimation and square-root CA excel in access-dominated cases, but in dense multi-UE scenarios, ACK-based CN estimates and AKF corrections are biased by cross-UE contention or non-access bottlenecks. Controller misattribution of transient measurements leads to quick-start decisions causing buffer blooms on slower paths and higher PLR/OFO. PHACC mitigates this by fusing physical estimation with GPASP’s latent cross-flow knowledge and per-subflow historical windows, making its quick-start aggressive. Therefore, it exhibits poor performance in terms of PLR and OFO rate.

      \textit{4) OFO degree comparison:} Fig.\ref{fig:p2}\subref{fig:p2d} presents OFO degree distributions: PHACC achieves the lowest median and tightest Interquartile Range (IQR), while MACO exhibits the largest variance and highest OFO rate (\(\simeq18.8\%\)). Ablation shows removing GPASP significantly increases OFO rate and OFO degree, confirming accurate scheduler priors help CC avoid window-driven reordering. This is because aggressive yet uninformed cwnd increases (e.g., quick-start) push packets to temporarily faster paths, with later arrivals on slower paths causing severe receiver reordering. PHACC avoids this via EDBSS-tempered early growth, SNR/load-aware pacing adaptation across subflows, and window-scheduler prediction synchronization to steer bulk transfers away from degrading paths, reducing OFO degree and rate.

  \subsection{NNPE performance verification}

    We validate the NNPE) designed to mitigate inference-induced decision staleness and deliver low-latency, suboptimal scheduling recommendations when full NN inference is computationally prohibitive. Experiments compare the full GPASP agent (with NNPE enabled) against a NNPE-disabled configuration (``w/o NNPE''), with all other components unchanged. Key evaluation metrics include throughput convergence, delay/jitter/PDR, and decision latency (CPU and wall-clock per decision).

    {\begingroup
    \let\origsize\normalsize
    \renewcommand{\normalsize}{\scriptsize}%
    \begin{figure}[!ht]
      \centering
      \subfloat[Comprehensive performance]{\includegraphics[width=0.24\textwidth]{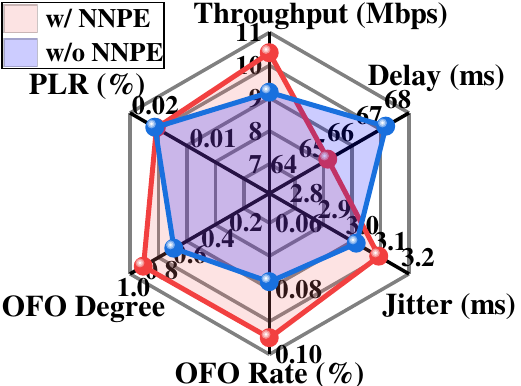}%
      \label{fig:p3a}}
      \subfloat[Runtime Efficiency]{\includegraphics[width=0.24\textwidth]{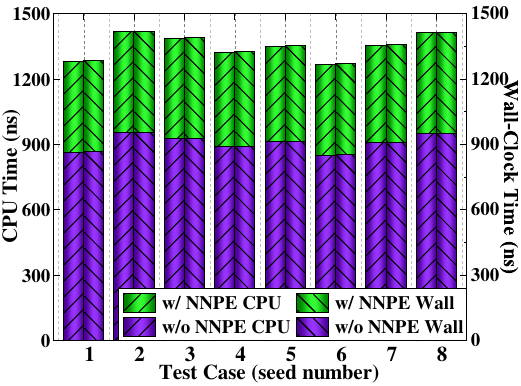}%
      \label{fig:p3b}}
      \caption{Comparison of NNPE ablation.}
      \label{fig:p3}
    \end{figure}
    \endgroup}

    Fig.\ref{fig:p3}\subref{fig:p3a} and Fig.\ref{fig:p3}\subref{fig:p3b} show that NNPE preserves throughput and QoS relative to the full-online-inference baseline. This is because NNPE ensures decisions meet the required control frequency: while suboptimal to the full policy in quasi-static environments, NNPE outperforms in highly time-varying regimes by avoiding losses from stale decisions, embodying a timeliness-vs.-optimality trade-off favoring timeliness in SAGIN operations. Fig. 2(b) further compares per-test-case CPU and wall-clock for runs with/without NNPE, demonstrating that NNPE significantly improves computational inference performance by approximately 100\%. This improvement decreases decision staleness in fast-changing slots, thereby lowering inference-induced mis-schedules.

    Instead of executing a full deep policy forward pass at each scheduling slot, NNPE computes a linearized preference-weight vector $\hat\theta_n$ from lightweight statistics (ACK success/loss $c_i^n$ and response times $\tilde t_i^n$ via the closed-form estimator in \eqref{eq:estimator}. This replaces costly nonlinear inference with a lightweight linear algebra update (e.g., recursive least-squares or low-dimensional incremental solvers) implementable incrementally, substantially reducing per-decision wall-clock time while generating actionable scheduling hints.

    NNPE relies on recent ACK statistics, requiring minimal samples to generate a stable $\hat\theta_n$, yet may be biased under sparse ACKs (very low sending rate) or highly bursty losses (e.g., sustained deep fades). To address this, we: (i) smooth $\hat\theta_n$ via exponential moving average (EMA); (ii) revert to conservative scheduling heuristics when sample counts drop below a threshold; and (iii) employ the Resilient Hierarchical Reward Monitor (RHRM) to detect persistent NNPE degradation, triggering lightweight parameter updates or full policy retraining as needed. These safeguards are integrated into the evaluation, preventing NNPE from destabilizing the system under pathological conditions.

    NNPE yields maximal benefits under three conditions: (i) the environment has short-term predictability (rendering preference estimates meaningful); (ii) full policy inference latency is non-negligible relative to control slot length; and (iii) computational resources are constrained (e.g., edge devices, UAV on-board controllers). Full policy inference may be preferable for deployments with abundant compute and slowly varying environments; otherwise, NNPE serves as a pragmatic, low-latency alternative.

  \section{Conclusion}

    In this paper, we conducted the first systematic investigation into the optimization of MPQUIC tailored for multi-access UAV-assisted SAGINs. Recognizing the intrinsic coupling and frequent mismatch between aggressive congestion control and conservative scheduling, we developed the GPR Hierarchical Synergistic Framework to maximize overall system throughput while minimizing OFO degrees. By integrating the GPASP module, our design successfully navigates the partial observability and high-dimensional noise inherent in multi-user environments. The introduction of the PHACC algorithm addresses severe throughput fluctuations by granting the system a priori handover awareness and enabling proactive window adjustments through multi-metric loss differentiation. Moreover, the NNPE algorithm effectively mitigates neural network inference latency to ensure synchronized packet dispatching , while the RHRM mechanism guarantees long-term robustness against environmental stochasticity and distributional shifts. Comprehensive evaluations on the extended ns-3 platform validate that our joint optimization framework achieves state-of-the-art transmission capacity, exceptional delay-jitter tradeoffs, and highly stable network performance. For future work, we will extend this synergistic framework to complex multi-orbit SAT architectures, refine the optimization model to integrate UAV energy constraints and matched topological rules, and further generalize the framework for wide-ranging 6G deployments.


  \bibliographystyle{unsrt}
  \bibliography{Reference}

  \vfill

\end{document}